\documentclass[journal,draftclsnofoot,onecolumn]{IEEEtran}
\usepackage{cite}
\usepackage{multicol}
\usepackage{multirow}
\usepackage{amsfonts}
\usepackage{amsmath}
\usepackage{amssymb}
\usepackage[english]{babel}
\newcommand{\be}{\begin{equation}}
\newcommand{\ee}{\end{equation}}
\newcommand{\bea}{\begin{eqnarray}}
\newcommand{\eea}{\end{eqnarray}}
\newcommand{\beax}{\begin{eqnarray*}}
\newcommand{\eeax}{\end{eqnarray*}}
\newcommand{\E}{\mathop{\mathbb E}}
\newcommand{\PR}{\mathop{\mathbb P}}

\newtheorem{teorema}{{\bf Theorem}}
\newtheorem{lemma}{{\bf Lemma}}
\newtheorem{corollario}{{\bf Corollary}}
\newtheorem{property}{{\bf Property }}

\begin{document}

\title{A formal proof of the optimal frame setting for Dynamic-Frame Aloha with known population size}

 \author{Luca~Barletta, Flaminio~Borgonovo, and~Matteo~Cesana
 %\IEEEauthorblockN{Luca Barletta, Flaminio Borgonovo, Matteo Cesana}
\thanks{L. Barletta was formerly with the Dipartimento di Elettronica,
Informazione e Bioingegneria, Politecnico
di Milano, Italy. Now he is with the Institute for Advanced Study, Technische
Universit\"at M\"unchen, Germany. (e-mail: luca.barletta@tum.de)}
\thanks{F. Borgonovo and M. Cesana are with the Dipartimento di Elettronica,
Informazione e Bioingegneria, Politecnico
di Milano, Italy (e-mail: borgonov@elet.polimi.it; cesana@elet.polimi.it).}}
% \IEEEauthorblockA{Dipartimento di Elettronica e informazione\\
% Politecnico di Milano\\
% Piazza L. da Vinci 32, \\
% 20133 Milan, Italy\\
% Email: \{barletta, borgonov, cesana\}@elet.polimi.it}

\maketitle

\begin{abstract}
In Dynamic-Frame Aloha subsequent frame lengths must be optimally chosen to maximize throughput.
When the initial population size ${\cal N}$ is known, numerical evaluations show that the maximum
efficiency is achieved by setting the frame length equal to the backlog size at each subsequent
frame; however, at best of our knowledge, a formal proof of this result is still missing, and is
provided here. As byproduct, we also prove that the asymptotic efficiency in the optimal case is
$e^{-1}$, provide tight upper and lower bounds for the length of the entire transmission period and show
that its asymptotic behaviour is $\sim ne-\zeta \ln (n)$, with $\zeta=-0.5/\ln(1-e^{-1})$.
\end{abstract}


%\IEEEpeerreviewmaketitle

\section{Introduction}
\label{sec:intro} Collision resolution protocols have played a fundamental role in communication
systems starting with the appearance of the Aloha protocol \cite{aloha1, roberts, sidi} back in 1970. Since then, a variety of such protocols have been proposed and
have influenced satellite, radio and local area networks, being nowadays applied
also to radio frequency identification (RFID) systems \cite{zhu2, rfid2}. In RFID systems a
reader interrogates a set of tags in order to identify each of them \cite{rfid2}. Collisions may
occur among the responses of tags and collision resolution protocols are used to arbitrate
collisions so that all tags can be finally identified. In this environment, the number of tags to
be identified ${\cal N}$ is not a random variable as it happens in multiple access systems, but is
a constant $n$, either known or unknown; nevertheless, the collision resolution problem is quite
similar in both environments and RFID protocols often represent a straightforward derivation of those proposed
for multiple access.

Among the different protocols envisaged in past years, Dynamic Frame Aloha (DF-Aloha) is the most
popular in RFID \cite{ISO, EPC}. In Frame-Aloha (F-Aloha) time is divided into time slots
equal to a packet transmission time, slots are grouped into frames, and a tag is allowed to
transmit only a single packet per frame in a randomly chosen slot. In the first frame all tags
transmit, but only a part of them avoid collisions with other transmissions and get through. The
remaining ones, referred to as the backlog, re-transmit in the subsequent frames until all of them
are successful. Although some versions allow the restart of a new frame at any slot, should this be convenient, here we deal with the original one, where the frame is explored in its entirety.

Unfortunately F-Aloha, like other protocols of the Aloha family \cite{gelembe, borovkov}, is
intrinsically unstable and its throughput is very small unless some stabilizing control is used. A
way to do this is to dynamically adapt the frame length $r$ according to the backlog size $n$,
hence the name Dynamic Frame Aloha (DF-Aloha). This strategy has been proposed for the first time
in \cite{schoute}, in the field of satellite communications, where the author proposes to set the
frame length exactly equal to a backlog estimate $\hat n$. The reason for adopting this strategy
is that the throughput in a slot of a frame of length $r$:
 \be \frac{n}{r}\left(1-\frac{1}{r}\right)^{n-1}, \ee
is maximized for $r=n$.

As a matter of fact, the performance figure to be optimized in choosing the size of each frame is
the overall efficiency
 \be \eta=\frac{E[{\cal N}]}{E[{\cal L}]}, \ee
where ${\cal N}$ is the original tag population size and ${\cal L}$ is the average  length of the identification period (IP), i.e., the average  number of slots needed to successfully transmit all the ${\cal N}$ tags. In
RFID systems ${\cal N}$ is usually a constant $n$ and, therefore, the efficiency is maximized by
minimizing $L(n)=E[{\cal L}]$.

A recursive formula is given \cite{schoute} for the calculation of $L(n)$. By applying this formula with known $n$, we can numerically show that the strategy that sets $r=n$ at each frame provides the shortest $L(n)$ for any value of $n$ attempted.
However, up to now, to the knowledge of authors, none has provided a theoretical verification of
the result.

In practical RFID applications $n$ is usually unknown; however, in order to meet optimality
conditions, it is usually replaced by an estimate $\hat n$ based on the observation of outcomes in
a frame or in the entire history (see for example
\cite{zhu2, vogt, flor2, tcom13}). Quite often setting $r=\hat n$ has been assumed,
never really discussing the optimal strategy when $n$ is unknown and an estimate is needed, with the notable exception
of \cite{zhu}. In this paper the authors have pointed out the non-optimality of the above setting,
and suggest a procedure to numerically find the best frame-length choice when the initial backlog
size $n$ is known in distribution. This procedure, when applied to known $n$, provides the
recursive formula cited above for $L(n)$, which is still solved only numerically.

In this paper we present an analysis of DF-Aloha with known backlog size $n$, that definitely proves that local optimization, i.e., maximizing the throughput/efficiency in each frame (i.e., setting $r=n$ at each frame), also maximizes the overall efficiency. We rigorously prove that the optimal asymptotic efficiency is $e^{-1}$, further providing tight upper
and lower bounds for $L(n)$, and showing that its asymptotic behaviour is $\sim ne-\gamma \ln (n)$,
with $\gamma=-0.5/\ln(1-e^{-1})$.

The Proof starts with providing, in Theorem \ref{th:elle1}, some general properties of $L(n)$ with
strategy $r=n$ at each frame, i.e., the strategy we will prove to be optimal. In particular, we
show that $L(n)$ is an increasing function upper bounded by $ne$. We then demonstrate two lemmas
about upper and lower bounds for the derivative of $\varepsilon(n)=L(n)-ne$. Subsequent lemmas provide lower and
upper bounds to error $\varepsilon(n)$, and together provides Theorem \ref{th:bounds} that in turn
gives the error's asymptotic behaviour. Finally, from all the preceding results, we are able to prove main Theorem \ref{th:ottimizz}, that confirms the optimality of the cited strategy. The key lemmas, together with the final theorem, make
use of some properties of statistical dominance of the first order \cite{fishburn}. To this end,
Appendix A extends a known result of statistical dominance, and provides a lemma where the
distributions of collided tags in frame $(n,r)$, for different $r$ and $n$, are ranked in terms of
such a statistical dominance. Often the proofs are analytically valid starting from a population
size $n=n_0$ somewhat greater than zero, and implying numerical verification up to $n_0$. To this
purpose, in Appendix B, we provide the distribution of the number of successes in a frame $(n,r)$.
Additional proofs of some properties used in theorems and lemmas are also given.

\section{Analysis} \label{sec:proof}

Let $n$ be the number of tags to be identified and $L(n,r_n)$ the average length of the
identification period, where we have made explicit its dependence on $r_n$, the length of the
frame with $n$ tags. The latter can be expressed as \cite{schoute}
\begin{equation}
L(n, r_n)=r_n+\sum_{s=0}^{m} p_{n, r_n}(s) L(n-s, r_{n-s}), \qquad n\ge 2, \label{eq:x2}
\end{equation}
where $m=\min \{n-2, r_n-1\}$, and $ p_{n, r_n}$ is the
probability distribution of the number of successes ${\cal S}_{n,r_n}$ in the first frame, of length $r_n$.
Making the term $L(n, r_n)$ explicit yields
\begin{equation}
L(n, r_n)=\frac{\displaystyle r_n+\sum_{s=1}^{m} p_{n, r_n}(s) L(n-s, r_{n-s})}
{1-p_{n, r_n}(0)},\qquad n\ge 2. \label{eq:x3}
\end{equation}
If the sequence $\{r_n\}$ is known, then (\ref{eq:x3}) can be used recursively to get the sequence
$L(n,r_n)$ starting from $L(0, r_0)=L(1, r_0)=r_0$.

A recursive expression of $p_{n, r_n}(s,c)$, the probability of having $s$ successes and $c$
collided slots in the first frame, is given in \cite{schoute}.
A closed form expression for $p_{n, r_n}(s)$, given in Appendix \ref{sec:appendix_B}, can be derived from formulas in \cite{feller}.

Let now call "Selected Strategy" the one that assumes $r_n=n$ at all frames. Later in the paper
(Theorem \ref{th:ottimizz}) we show that the Selected Strategy is indeed the optimum strategy. In
the remainder of the paper, for the sake of compactness, when $r=n$ we use a single subscript in
the notation, e.g. ${\cal S}_{n}$ in place of the more general ${\cal S}_{n,n}$. In the
analysis that follows we also make use of some properties of random variable (RV) $\cal{S}$, and the related
${\cal R}=n-{\cal S}$, that are listed in Appendix \ref{sec:appendix_B}.

Expressions (\ref{eq:x2}) and (\ref{eq:x3}), with the Selected Strategy can be rewritten as
follows, being the dependence on $r_n$ omitted:
\begin{align}
L(n) &= n+\sum_{i=1}^{n} \pi_{n}(i) L(i), \qquad n\ge 2, \label{eq:x5} \\
L(n)&=\frac{\displaystyle n+\sum_{i=1}^{n-1} \pi_{n}(i) L(i)}{1-\pi_{n}(n)}, \qquad n\ge 2,
\label{eq:x5a}
\end{align}
where $\pi_{n,r}(i)= p_{n,r}(n-i)$, for $i=0,\ldots,n$, is the probability distribution of RV ${\cal R}=n-{\cal S}$, the number of collided tags out of the initial $n$. Note that $\pi_{n,r}(1)=0$ for any pair $(n,r)$,
therefore the summation in (\ref{eq:x5}) can be started from $i=2$.

We now prove the following:
\begin{teorema}\label{th:elle1}
with the Selected Strategy the average identification period in identifying $n$ tags, $L(n)$,
presents the following properties:
 \begin{enumerate}
 \item [(a)] $L(n)$ is an increasing function of $n$,
 \item [(b)] $ L(n)< ne$.
 \end{enumerate}
 \end{teorema}
\begin{IEEEproof}
 (a) For $n=1$ the thesis holds because $L(1)=1$ and $L(2)=4$. We assume that
$L(i)> L(i-1)$ for $i \le n$ and show that it holds also for $i=n+1$.

We can easily lower bound the difference between (\ref{eq:x5a}), evaluated in $n+1$, and (\ref{eq:x5}) as follows
\be L(n+1)-L(n) >1 + \sum_{i=2}^{n}(\pi_{n+1}'(i)-\pi_n(i))L(i),\label{eq:th25}\ee
where 
\be\pi_{n}'(i)=\frac{\pi_{n}(i)}{1-\pi_{n}(n)}, \qquad 0\le i \le n-1. \label{eq:norm} \ee

%Relation (\ref{eq:x5a}) can be written, for $n+1$, as
% \be L(n+1)=\frac{n+1}{1-\pi_{n+1}(n+1)} + \sum_{i=2}^{n}\pi^*_{n+1}(i)
%L(i)\ge n+1 + \sum_{i=2}^{n}\pi^*_{n+1}(i) L(i),\label{eq:th23}\ee where we have set
%\be\pi^*_{n}(i)=\frac{\pi_{n}(i)}{1-\pi_{n}(n)}, \qquad 0\le i \le n-1. \label{eq:norm} \ee
%
%By subtracting (\ref{eq:x5}) to (\ref{eq:th23}) we get
% \be L(n+1)-L(n) >1 + \sum_{i=2}^{n}(\pi^*_{n+1}(i)-\pi_n(i))L(i).\label{eq:th25}\ee
Since by Lemma \ref{th:dominance1} distribution $\pi_{n+1}$ statistically dominates $\pi_{n}$, it
is easy to prove that also $\pi'_{n+1}$ statistically dominates $\pi_{n}$. Therefore, owing to the
fact that $L(i), i\le n$, is an increasing function of $i$, by the basic property
(\ref{eq:app01}), the summation in (\ref{eq:th25}) can not be negative, and the thesis is proved.

(b) Let us assume that $L(i)< ie$ for $i \le n$ and show that it holds also for $i=n+1$ (it is
trivially $L(1)=1<e$). From (\ref{eq:x5a}), the assumption allows to write
\begin{align} L(n+1)&<
\frac{\displaystyle n+1+\sum_2^{n}\pi_{n+1}(i)ie}{1-\pi_{n+1}(n+1)}
=\frac{\displaystyle n+1+(R_{n+1}- (n+1)\pi_{n+1}(n+1))e }{1-\pi_{n+1}(n+1)},
\label{eq:th42a}\end{align}
 where $R_n=\E[{\cal R}_n]$. By Property \ref{th:properties}b of Appendix B, we have $ R_{n} < n(1-e^{-1})$, which, used in (\ref{eq:th42a}), finally provides
 \be L(n+1)<\frac{\displaystyle (n+1 - (n+1)\cdot\pi_{n+1}(n+1))e }{1-\pi_{n+1}(n+1)}=(n+1)e.
\ee
\end{IEEEproof}

Denoting by $\Delta f(n)= f(n)-f(n-1)$ the derivative of function $f(n)$, the next lemma provides
an upper bound to the derivative of error $ \varepsilon(n)=ne -L(n)$.

\begin{lemma}\label{th:min_Delta_errore}
given the function
 \be g(n)= \nu \ln(n) + \frac{\mu}{\sqrt n}, \label{eq: th a20} \ee
the following inequality holds for $\nu=1.5$ and $\mu=2$:
 \be \Delta \varepsilon(n)\le \Delta g(n),\qquad n\ge 2. \label{eq: th20} \ee
\end{lemma}
\begin{IEEEproof}
by (\ref{eq:x5}) we have
 \be \varepsilon(n)=ne -L_{n}= n(e-1) -R_n e +\sum_2^{n}\pi_{n}(i)\varepsilon(i), \qquad n\ge 2. \label{eq: err10} \ee
The general term of the difference sequence for $n \ge 2$ can be written as 
 \begin{align} \Delta \varepsilon(n+1)&= e-1
- (R_{n+1}-R_n) e + \sum_{i=2}^n \varepsilon(i) (\pi_{n+1}'(i) - \pi_n(i)) + \nonumber \\
+ & \frac{\pi_{n+1}(n+1)}{1-\pi_{n+1}(n+1)} ((n+1)(e-1)-R_{n+1}e) < \sum_{i=2}^{n}\varepsilon(i)(\pi'_{n+1}(i)-\pi_{n}(i))
 +\mathcal{O}(c^{n}), \label{eq:aw1}
\end{align}
where we have exploited inequality $R_{n+1}- R_n > (1-e^{-1}) $ derived in Property \ref{th:properties}c of Appendix B,
and $\mathcal{O}(c^{n})$ corresponds to fractional term on the left in (\ref{eq:aw1}) with $c=0.9157$ being derived in Property
\ref{lem:expo_pi}. If we now assume that
 \be \Delta \varepsilon(i) \le \Delta g(i), \qquad 2 \le i \le n \label{eq:w1}\ee
we show that
 \be \Delta \varepsilon(n+1)< \Delta g(n+1), \qquad n \ge 2, \ee
proving the theorem by induction. Since,
by Lemma \ref{th:dominance1}, distribution $\pi_{n+1}$ statistically dominates distribution
$\pi_{n}$ in the first order, it is easy to verify that also distribution $\pi'_{n+1}$
statistically dominates distribution $\pi_{n}$ in the first order. Therefore, by Lemma
\ref{th:dominance} of Appendix A and (\ref{eq:w1}),  we can write
 \be \overline{\varepsilon(i)} =\sum_{i=1}^{n} \varepsilon(i) (\pi'_{n+1}(i)-\pi_{n}(i))\le
\sum_{i=1}^{n}g(i)(\pi'_{n+1}(i)-\pi_{n}(i))= \overline{g(i)}=\nu
\overline{\ln(i)}+\overline{\mu/\sqrt i}. \label{eq:dd2} \ee

 In order to upper bound $\overline{g(i)}$ we focus on  term $\overline{\ln(i)}$, and make use again of Lemma
\ref{th:dominance} of Appendix A in the form provided by  Corollary \ref{th:dominancex}, by writing
 \be \overline{\ln(i)} = \sum_{i=1}^{n}\ln(i)(\pi'_{n+1}(i)-\pi_{n}(i))
\le \sum_{i=1}^{n}g_1(i)(\pi'_{n+1}(i)-\pi_{n}(i)),
 \label{eq:kk2} \ee
where
 \be g_1(x)=\sum_{i=1}^{4}(-1)^{i+1}   \frac{(x-a)^i}{ia^i} + \frac{(x-a)^5}{5a^4}, \qquad \forall x \ge 1, \forall a \ge 1. \label{eq:rr2} \ee
% \be g_1(x)=\frac{x-a}{a} - \frac{(x-a)^2}{2a^2} + \frac{(x-a)^3}{3a^3}- \frac{(x-a)^4}{4a^4}
% + \frac{(x-a)^5}{5a^4}, \qquad \forall x \ge 1, \forall a \ge 1. \label{eq:rr2} \ee
and
 \be \frac{d \ln x}{dx} \le \frac{d g_1(x)}{dx} = \sum_{i=0}^{3}(-1)^{i}   \frac{(x-a)^i}{a^{i+1}} + \frac{(x-a)^4}{a^4}, \qquad \forall x \ge 1, \forall a \ge 1. \label{eq:ra1}
\ee
% \be \frac{d \ln x}{dx} \le \frac{d g_1(x)}{dx} = \frac{1}{a} - \frac{(x-a)}{a^2} + \frac{(x-a)^2}{a^3} -
%\frac{(x-a)^3}{a^4} + \frac{(x-a)^4}{a^4}, \qquad \forall x \ge 1, \forall a \ge 1 \label{eq:ra1}
%\ee

By choosing $a=R_n$in (\ref{eq:rr2}) and expliciting it in (\ref{eq:kk2}) the latter becomes 
 \begin{align} \overline{\ln(i)} &\le \sum_{i=1}^{4}(-1)^{i+1} \left(\frac{\E[({\cal R}'_{n+1}-R_n)^i]}{iR_n^i}-\frac{\E[({\cal R}_{n}-R_n)^i]}{iR_n^i}\right)+
 \frac{\E[({\cal R}'_{n+1}-R_n)^5]}{5R_n^4} - \frac{\E[({\cal R}_{n}-R_n)^5]}{5R_n^4} \label{eq:kk3}\\
& =n^{-1} +\frac{\left(6+3 e-e^2\right) }{2e (e-1)} n^{-2} + \mathcal{O}(n^{-3}), \label{eq:hh3}\end{align}
%\[\overline{\ln(i)} \le \frac{R'_{n+1}-R_n}{R_n} - \frac{E[({\cal R}'_{n+1}-R_n)^2]}{2R_n^2} +
%\frac{E[({\cal R}'_{n+1}-R_n)^3]}{3R_n^3} - \frac{E[({\cal R}'_{n+1}-R_n)^4]}{4R_n^4} +
%\frac{E[({\cal R}'_{n+1}-R_n)^5]}{5R_n^4} + \] \be \hspace{1cm} + \frac{E[({\cal
%R}_{n}-R_n)^2]}{2R_n^2}- \frac{E[({\cal R}_{n}-R_n)^3]}{3R_n^3} + \frac{E[({\cal
%R}_{n}-R_n)^4]}{4R_n^4} - \frac{E[({\cal R}_{n}-R_n)^5]}{5R_n^4}
% \label{eq:kk3} \ee
where in the last step moments of RVs ${\cal R}_{n+1}'$ and ${\cal R}_n$ have been computed as described in Property 5 of Appendix B,
and Taylor's expansions have been used. % the following asymptotic expansions
% \begin{align}
% \frac{R'_{n+1}-R_n}{R_n} &=n^{-1}+\frac{1}{2 (e-1)}n^{-2}
% +\mathcal{O}(n^{-3})
% \\
% -\frac{E[({\cal R}'_{n+1}-R_n)^2]}{2R_n^2}+ \frac{E[({\cal R}_{n}-R_n)^2]}{2R_n^2}&=
% -\frac{e}{2(e-1)}en^{-2}+ \mathcal{O}(n^{-3})
%\\
% \frac{E[({\cal R}'_{n+1}-R_n)3]}{3R_n^3}-\frac{E[({\cal R}_{n}-R_n)3]}{3R_n^3}&= \frac{1}{e-1}n^{-2}
% + \mathcal{O}(n^{-3})\\
% -\frac{E[({\cal R}'_{n+1}-R_n)^4]}{4R_n^4}+\frac{E[({\cal R}_{n}-R_n)^4]}{4R_n^4}&=
% \mathcal{O}(n^{-3})\\
% \frac{E[({\cal R}'_{n+1}-R_n)^5]}{5R_n^4}-\frac{E[({\cal R}_{n}-R_n)^5]}{5R_n^4}&=
% \frac{3}{e(e-1)}n^{-2} + \mathcal{O}(n^{-3})
% \end{align}
% and get
% \be\overline{\ln(i)} \le n^{-1} +\frac{\left(6+3 e-e^2\right) }{2e (e-1)} n^{-2} + \mathcal{O}(n^{-3}). \label{eq:hh3} \ee
 We repeat the same procedure for term $1/\sqrt n$ in $g(n)$. We have
\be \frac{d (1/\sqrt x)}{dx} \le - \frac{1}{2 \sqrt{a^3}}+ \frac{3(x-a)}{4 \sqrt{a^5}}, \qquad
\forall x \ge 1, \forall a \ge 1 \ee
 which implies
\be \overline{1/\sqrt i}=\sum_{i=1}^{n}(1/\sqrt i)(\pi'_{n+1}(i)-\pi_{n}(i)) \le
\sum_{i=1}^{n}g_2(i)(\pi'_{n+1}(i)-\pi_{n}(i)),
 \label{eq:kk22} \ee
being
 \be g_2(x)=-\frac{x-a}{2 \sqrt{a^3}} + \frac{3(x-a)^2}{8 \sqrt{a^5}}, \qquad \forall x \ge
1, \forall a \ge 1. \ee

Again, choosing $a=R_n$, inequality (\ref{eq:kk22}) becomes
 \begin{align} \overline{1/\sqrt i} &\le -\frac{R'_{n+1}-R_n}{2 \sqrt{R_n^3}} + \frac{3\E[({\cal
R}'_{n+1}-R_n)^2]}{8 \sqrt{R_n^5}}- \frac{3\E[({\cal R}_{n}-R_n)^2]}{8 \sqrt{R_n^5}}
 \label{eq:kr22}\\
&=\frac{1}{2 \sqrt{1-e^{-1}}}\ n^{-3/2} + \mathcal{O}(n^{-5/2}), \label{eq:hh33} \end{align}
%and replacing the corresponding asymptotic expansions one gets
%\[ \frac{R'_{n+1}-R_n}{2 \sqrt{R_n^3}} =
%\frac{1}{2 \sqrt{1-e^{-1}}}\ n^{-3/2}+\mathcal{O}(n^{-5/2})\]
%\[ \frac{3E[({\cal
%R}'_{n+1}-R_n)^2]}{8 \sqrt{R_n^5}}- \frac{3E[({\cal R}_{n}-R_n)^2]}{8 \sqrt{R_n^5}}=\mathcal{O}(n^{-5/2})\]
% we get
% \be\overline{1/\sqrt i} \le \frac{1}{2 \sqrt{1-e^{-1}}}\ n^{-3/2} + \mathcal{O}(n^{-5/2}), \label{eq:hh33} \ee
and putting together (\ref{eq:hh3}) and (\ref{eq:hh33}) we finally get \be \overline{g(i)}=
\nu \overline{\ln(i)}+\overline{\mu/\sqrt i} \le \nu n^{-1}+ \mu \frac{1}{2 \sqrt{1-e^{-1}}}\
n^{-3/2} +\mathcal{O}(n^{-2}). \label{eq:hhh33} \ee

 From (\ref{eq:aw1})
and (\ref{eq:dd2}), we see that, to prove the thesis, we must have
 \begin{align*} \overline{g(i)} + \mathcal{O}(c^{n}) \le \Delta g(n+1)&= \nu(\ln(n+1)-\ln(n)) +
\mu\left(\frac{1}{\sqrt{n+1}}-\frac{1}{\sqrt n} \right) \\
&=\nu \left( n^{-1}
 -\frac{1}{2}n^{-2}+\frac{1}{3}n^{-3}+\ldots \right) -\frac{\mu}{2} n^{-3/2}+ \dots.\end{align*}
%and taking the
% expansion of the right hand terms above the condition becomes
%  \be \overline{g(i)}+\mathcal{O}(c^{n}) \le \Delta g(n+1)= \nu \left( n^{-1}
% -\frac{1}{2}n^{-2}+\frac{1}{3}n^{-3}+\ldots \right) -\frac{\mu}{2} n^{-3/2}+ \dots. \ee
Substituting inequality (\ref{eq:hhh33}), term $n^{-1}$ simplifies and, to prove the thesis, the
final condition becomes
 \be \frac{\mu}{2 \sqrt{1-e^{-1}}}\ n^{-3/2} \le -\frac{\mu}{2} n^{-3/2}+ \mathcal{O}(n^{-2}) \label{eq:
th2c}. \ee
Disregarding the asymptotic term, the above inequality is always true for any $\mu$. This means
that there exists an $n_0$ such that for any $n \ge n_0$ (\ref{eq: th2c}) is satisfied.
%For example, with $\nu=1.5$ and $\mu=2$ we find $n_1=xxx$.
 Then, we numerically verify that (\ref{eq:
th20}) holds up to $n_0$, and the lemma is proved.
 \end{IEEEproof}
Note that, by proving (\ref{eq: th20}), the above Lemma proves also the following
\begin{corollario} \label{th:corol1} if distribution $\pi_X$ statistically dominates distribution
$\pi_Y$, and $g(n)$ is the function (\ref{eq: th a20}), then we have
 \be \sum_{i=1}^{n} \varepsilon(i) (\pi_X(i)-\pi_Y(i))\le
\sum_{i=1}^{n}g(i)(\pi_X(i)-\pi_Y(i)). \label{eq:ddx2} \ee
\end{corollario}

\begin{corollario} \label{th:corol2} if distribution $\pi_X$ statistically dominates distribution
$\pi_Y$, $g(n)$ is the function (\ref{eq: th a20}), and $l(n)=\nu \ln(n)$, $\nu$ as in (\ref{eq: th a20}), then we have
 \be \sum_{i=1}^{n} \varepsilon(i) (\pi_X(i)-\pi_Y(i))\le
\sum_{i=1}^{n}l(i)(\pi_X(i)-\pi_Y(i)), \label{eq:ddz2} \ee
\end{corollario}
where the thesis comes from Corollary \ref{th:corol1} and Lemma \ref{th:dominance}, being $\Delta l(n) \ge \Delta g(n)$.

 The next lemma provides a lower bound to $\Delta\varepsilon(n)$.
\begin{lemma} \label{th:max_Delta_errore} given the function
 \begin{align*} g(n)&= \nu \ln(n)\qquad n \ge 3 \\
 g(2)&= -0.5, \end{align*}
for $\nu=1$ the following inequality holds
 \be \Delta \varepsilon(n)\ge \Delta g(n),\qquad
n\ge 2. \label{eq: thb20} \ee
 \end{lemma}
\begin{IEEEproof} the proof proceeds exactly as in Lemma \ref{th:min_Delta_errore}, where now we set $g(n)=\nu
\ln(n)$. We exploit again (\ref{eq:aw1}),  carrying also
the infinitesimal terms of expansion in Property \ref{th:properties}c of Appendix B. This yields
 \be \Delta \varepsilon(n+1) = -\frac{7}{24n^2}+ \sum_{i=2}^{n} \varepsilon(i)(\pi'_{n+1}(i)-\pi_{n}(i))
+ \mathcal{O}(n^{-3}) , \qquad n\ge 2. \label{eq:w65} \ee
We proceed by induction as in Lemma \ref{th:min_Delta_errore}, where now (\ref{eq:dd2})
becomes $\overline{\varepsilon(i)}\ge \overline{g(i)}$.
% \be \overline{\varepsilon} =\sum_{i=1}^{n} \varepsilon(i) (\pi'_{n+1}(i)-\pi_{n}(i))\ge
%\sum_{i=1}^{n}g(i)(\pi'_{n+1}(i)-\pi_{n}(i))= \overline{g}. \label{eq:ddb2} \ee
Here the value
$g(2)$ has been chosen so as to satisfy $g(2)<\varepsilon(2)\approx -0.2817$.
 In order to lower bound $\overline{g(i)}$, relations (\ref{eq:rr2}) and (\ref{eq:ra1})  are respectively replaced by
\be g_1(x)=\sum_{i=1}^4 (-1)^{i+1} \frac{(x-a)^i}{ia^i},
\qquad \forall x \ge 1, \forall a \ge 1 \label{eq:krr2} \ee
 \be \frac{d \ln x}{dx} \ge \sum_{i=0}^3 (-1)^{i} \frac{(x-a)^i}{a^{i+1}}, \qquad \forall x \ge 1, \forall a \ge 1. \label{eq:ra2} \ee
% \be g_1(x)=\frac{x-a}{a} - \frac{(x-a)^2}{2a^2} + \frac{(x-a)^3}{3a^3}- \frac{(x-a)^4}{4a^4},
% \qquad \forall x \ge 1, \forall a \ge 1 \label{eq:krr2} \ee
%  \be \frac{d \ln x}{dx} \ge \frac{1}{a} - \frac{(x-a)}{a^2} + \frac{(x-a)^2}{a^3} -
% \frac{(x-a)^3}{a^4}, \qquad \forall x \ge 1, \forall a \ge 1. \label{eq:ra2} \ee
Using the same series expansions we get the corresponding of (\ref{eq:hh3}) as
 \be\overline{\ln(i)} \ge n^{-1}+ \frac{3-e}{2(e-1)} n^{-2} + \mathcal{O}(n^{-3}). \label{eq:hh44} \ee

From (\ref{eq:w65}), to prove the thesis $\Delta\varepsilon(n+1) \ge \Delta g(n+1)$, we must have
 \be -\frac{7}{24}n^{-2}+ \nu \overline{\ln(i)} +\mathcal{O}(n^{-3}) > \Delta g(n+1)= \nu
\left( n^{-1} -\frac{1}{2}n^{-2}+\frac{1}{3}n^{-3}+\ldots \right), \label{eq: th33c}\ee
and using (\ref{eq:hh44}) term $n^{-1}$ simplifies, leading to condition
 \be \frac{7}{24} n^{-2} \le \nu\frac{1}{e-1} n^{-2}+
\mathcal{O}(n^{-3}). \label{eq: th2cbis} \ee
If we disregard term $\mathcal{O}(n^{-3})$, for any $\nu \ge 0.502$ the above inequality is always
verified. This means that there is an $n_1(\nu)$ such that for $n\ge n_1(\nu)$ (\ref{eq: th2cbis}) is always verified,
and the lemma is proved if the thesis is shown to hold numerically up to $n_1(\nu)$. This is the
case, for example, with $\nu=1$, and the thesis is proved.
\end{IEEEproof}

Now we proceed to get bounds to error $\varepsilon(n)$ and to provide its asymptotic behaviour.

\begin{lemma}\label{th:minorazione_errore} the following inequality holds:
 \be \varepsilon(n) > f(n) =\zeta\ln(n)+ \frac{\lambda}{n}, \qquad n\ge 2, \label{eq:kr1}\ee
where $\lambda=1.2 $ and
 \be \zeta= -\frac{\displaystyle 0.5}{\ln(1-e^{-1})}=1.0900... \label{eq: kr2}\ee

\end{lemma}
\begin{IEEEproof} we start from relation (\ref{eq: err10})
that, using Property \ref{th:properties}b of Appendix B, $R_n/n= 1-e^{-1}-\xi_n$,
 becomes
 \[ \varepsilon(n)= n e\xi_n +\sum_2^{n}\pi_{n}(i)\varepsilon(i), \]
 and, solving for $\varepsilon(n)$, we get
 \be \varepsilon(n)= \frac{\displaystyle ne\xi_n +
 \sum_2^{n-1}\pi_{n}(i)\varepsilon(i)}{1-\pi_{n}(n)}. \label{eq: kerr3}
 \ee

In the following, we assume that inequality (\ref{eq:kr1}) is verified up to $n-1 \ge 2$, and show
that it is also satisfied for $n$, proving the theorem by induction. This assumption,
 applied to (\ref{eq: kerr3}), implies
 \be \varepsilon(n)> \ ne\xi_n + \sum_2^{n-1}\pi_{n}(i)f(i)= ne\xi_n +
 \sum_2^{n}\pi_{n}(i)f(i)- \pi_{n}(n)f(n). \label{eq: ker3}
 \ee
Similarly to preceding Lemmas we use inequalities \be \ln(x) \ge \ln(a) +\frac{(x-a)}{a} -
\frac{(x-a)^2}{2a^2}+ \frac{(x-a)^3}{3a^3}-\frac{(x-a)^4}{a^{3.5}}, \qquad \forall x \ge 1,
\forall a \ge 1, \label{eq:xra1} \ee
 \be \frac{1}{x} \ge \frac{1}{a} - \frac{(x-a)}{a^2},
 \qquad \forall x \ge 1, \forall a \ge 1, \label{eq:xra2} \ee
evaluated at $a=R_n$, which provide
 \be \overline{f(i)}=\sum_2^{n}\pi_{n}(i)f(i) \ge \zeta \left(
\ln(R_n)- \frac{\E[({\cal R}_{n}-R_n)^2]}{2R_n^2}+
 \frac{\E[({\cal R}_{n}-R_n)^3]}{3R_n^3} - \frac{\E[({\cal R}_{n}-R_n)^4]}{R_n^{3.5}}\right)
+ \frac{\lambda}{R_n}. \label{eq:exp12} \ee
 Using Property \ref{th:properties}b of Appendix B, and substituting Taylor's expansions,
% \begin{align}
% \ln(R_n)&=\ln(1-e^{-1})+\ln(n) -\frac{1}{2 (e-1)}n^{-1} + \mathcal{O}(n^{-2}), \label{eq:exlog}\\
% -\frac{E[({\cal R}_{n}-R_n)^2]}{2R_n^2}&=
%  -\frac{1}{2(e-1)}n^{-1}-\frac{e+1}{4(e-1)^2}n^{-2} + \mathcal{O}(n^{-3})
% \\
%  \frac{E[({\cal R}_{n}-R_n)^3]}{3R_n^3}&= -\frac{e-2}{3(e-1)^2}n^{-2}
%  + \mathcal{O}(n^{-3})\\
% -\frac{E[({\cal R}_{n}-R_n)^4]}{R_n^{3.5}}&= .... + \mathcal{O}(n^{-3})
%  \end{align}
(\ref{eq:exp12}) provides
 \be \overline{f(i)} \ge \zeta \left( \ln(1-e^{-1})+\ln(n) -\frac{1}{
(e-1)}n^{-1} \right) + \frac{\lambda e}{(e-1)n+1/2}
 + \mathcal{O}(n^{-2}). \label{eq:effe}\ee
From inequality (\ref{eq: ker3}), and using the above,  the thesis is true if the following holds
 \[ \varepsilon(n)> \ ne\xi_n +\zeta\ln(1-e^{-1})+\zeta\ln(n) - \frac{\zeta}{e-1}n^{-1}+
 \frac{\lambda e}{e-1}n^{-1} +
\mathcal{O}(n^{-2}) \ge \zeta\ln(n)+ \lambda n^{-1}. \]
 or \be ne\xi_n +\zeta\ln(1-e^{-1}) - \frac{\zeta}{e-1}n^{-1}+
 \frac{\lambda}{e-1}n^{-1} + \mathcal{O}(n^{-2}) \ge 0. \label{eq: mer3}
 \ee
Substituting expansion $ ne\xi_n = \frac{1}{2} +\frac{7}{24 n} + \mathcal{O}(n^{-2})$
 and $\zeta$, condition (\ref{eq: mer3}) becomes
\be \left( \frac{7}{24} - \frac{\zeta}{e-1}+ \frac{\lambda}{e-1}\right)n^{-1} + \mathcal{O}(n^{-2}) \ge
0. \label{eq: kker6} \ee
Disregarding term $\mathcal{O}(n^{-2})$, the above is true for $\lambda>\zeta-(7/24)(e-1)\approx
0.6$.  This means that, in this case, there is some $n_0(\lambda)$ such that for all $n\ge n_0$
(\ref{eq: kker6}) is true. The thesis is then proved by showing that (\ref{eq:kr1}) numerically
holds up to $n_0$. This happens for $\lambda=1.2$. Since we are dealing with a lower bound, we are
interested in taking $\lambda$ as large as possible. However, we have found that as $\lambda$
increases beyond $1.35$ (\ref{eq:kr1}) does not hold from $n=2$ up, and the lemma can not be
proved.
\end{IEEEproof}

\begin{lemma}\label{th:maggiorazione_errore} the following inequality holds:
 \be \varepsilon(n) < g(n) = \zeta\ln(n)+K, \qquad n\ge 2, \label{eq:1}\ee
where
 \be \zeta= -\frac{\displaystyle 0.5}{\ln(1-e^{-1})}=1.0900.., \label{eq: r2}\ee
 and $K=0.7$.
\end{lemma}
\begin{IEEEproof} the proof proceeds exactly as in Lemma \ref{th:minorazione_errore}, where term $\lambda/n$ is
replaced by the constant $K$. We assume that inequality (\ref{eq:1}) is verified up to $n-1 \ge 2$,
and show that it is also satisfied for $n$, proving the theorem by induction. The corresponding of (\ref{eq:
ker3}) is \begin{align} \varepsilon(n)&<
 \frac{1}{1-\pi_{n}(n)} \left(ne\xi_n +
 \sum_2^{n}\pi_{n}(i)g(i)- \pi_{n}(n)g(n) \right)=ne\xi_n +
 \sum_2^{n}\pi_{n}(i)g(i) + \mathcal{O}(c^{n}) \nonumber\\
&< ne\xi_n + \zeta\ln(R_n)+K + \mathcal{O}(c^{n}), \label{eq: er56}
 \end{align}
where in the last step we applied Jensen's inequality.
%  and get
%  \be \varepsilon(n)< ne\xi_n + \zeta\ln(R_n)+K + \mathcal{O}(c^{n}). \label{eq: er56}
%  \ee
 Again, using the expansion for $\ln(R_n)$ and Property \ref{th:properties}b of Appendix B, that provides
 $ ne\xi_n = \frac{1}{2} +\frac{7}{24}n^{-1} + \mathcal{O}(n^{-2})$, we have
 \be \varepsilon(n)< \frac{1}{2}
+\frac{7}{24}n^{-1} +\zeta\ln(1-e^{-1})+\zeta\ln(n) -\zeta \frac{1}{2 (e-1)}n^{-1} +K +
\mathcal{O}(n^{-2}).
 \label{eq: ker6}
 \ee

 The thesis holds if we show that $\varepsilon(n) < g(n)$, which, using the above and
 (\ref{eq: r2}), gives the condition
 \be \frac{7}{24}n^{-1} -\zeta \frac{1}{2 (e-1)}n^{-1} + \mathcal{O}(n^{-2}) < 0.
 \label{eq: ker6x}
 \ee
Disregarding term $\mathcal{O}(n^{-2})$, the inequality above is always true. This means that
there is some $n_0$ such that for all $n\ge n_0$ (\ref{eq: ker6x}) is true. The value of constant
$K$ has no effect on the above inequality; in fact, it is taken as the practical smaller value
that makes (\ref{eq:1}) true for $2 \le n\le n_0$. We have found that the thesis holds with $K=0.7$.
\end{IEEEproof}

Using the preceding lemmas we may conclude:
\begin{teorema}\label{th:bounds}
 \be \frac{1.2}{n} + \zeta\ln(n) \le \varepsilon(n) < 0.7
 + \zeta\ln(n), \qquad n\ge 2, \label{eq:1ww}\ee
 \be \varepsilon(n) \sim \zeta\ln(n). \ee
\end{teorema}

We now are in the position to prove the main theorem of this paper. Let $L(n,r) $ be the average
IP when $r$ is the length of the first frame, whereas for the remaining frames the Selected
Strategy is adopted.

\begin{teorema}\label{th:ottimizz} $L(n,r) $ is minimized by the strategy that at each frame sets the frame length
 $r$ equal to the backlog size $n$.
\end{teorema}
\begin{IEEEproof} we assume that the above strategy is used in all frames with backlog $i$, $2
\le i \le n-1$, and show that we have
\begin{align}
 \Delta L(n,k)&=L(n,n)-L(n,n+k)<0, \quad k \ge -(n-1), \quad k \ne 0, \label{eq:a3}
\end{align}
then the theorem is proved by induction starting from $n=2$.
From (\ref{eq:x5}) and (\ref{eq:x5a}) we have
 \be \Delta L(n,k)= n+\sum_{i=2}^{n}
\pi_{n,n}(i) L(i)-\frac{n+k+\sum_{i=2}^{n-1} \pi_{n,n+k}(i) L(i)}{1-\pi_{n,n+k}(n)}.
\label{eq:a14o}\ee
We now proceed by proving (\ref{eq:a3}) for the two cases, $k<0$ and $k>0$.

\textbf{Part Ia}. $k<0$, or $h=-k>0$. Since the range of $h$ depends on $n$, we further set  $h=\alpha n$, with $1/n \le \alpha \le (n-1)/n$.
Equation (\ref{eq:a14o}) can be expressed as \be \Delta L(n,-\alpha n)= n+\sum_{i=2}^{n} \pi_{n,n}(i)
L(i)-\frac{n(1-\alpha)+\sum_{i=2}^{n} \pi_{n,n(1-\alpha)}(i) L(i)- \pi_{n,n(1-\alpha)}(n)
L(n)}{1-\pi_{n,n(1-\alpha)}(n)}. \label{eq:a14w}\ee
By Theorem \ref{th:bounds} we use inequalities
\[ ie-\zeta\ln i-0.7\le L(i) \le ie - \zeta \ln(i), \]
%\[ L(i) \ge ie - \zeta \ln(i) - 0.7, \]
to bound (\ref{eq:a14w}) as follows
\begin{align} \Delta L(n,-\alpha n) & \le n+eR_{n,n} - \zeta \E[\ln {\cal R}_{n,n}] \nonumber\\
 &\quad-\frac{n(1-\alpha)+e R_{n,n(1-\alpha)} - \zeta \E[\ln {\cal R}_{n,n(1-\alpha)}]- 0.7 -\pi_{n,n(1-\alpha)}(n)(ne-\zeta \ln(n)) }{1-\pi_{n,n(1-\alpha)}(n)} \nonumber\\
 &= -n(e-1)+eR_{n,n} + \zeta( \ln(n)- \E[\ln {\cal R}_{n,n}]) \label{eq:akkk}\\
 &\quad-\frac{- n(e-1)+e R_{n,n(1-\alpha)}+\zeta ( \ln(n) - \E[\ln {\cal R}_{n,n(1-\alpha)}])-n\alpha- 0.7 }{1-\pi_{n,n(1-\alpha)}(n)}. \label{eq:a15k}\end{align}
% \begin{align} \Delta L(n,-\alpha n) & \le n+eR_{n,n} - \zeta \E[\ln {\cal R}_{n,n}] \\
%  &\quad-\frac{n(1-\alpha)+e R_{n,n(1-\alpha)} - \zeta \E[\ln {\cal R}_{n,n(1-\alpha)}]- 0.7 -\pi_{n,n(1-\alpha)}(n)(ne-\zeta \ln(n)) }{1-\pi_{n,n(1-\alpha)}(n)} \\
%  &= n+eR_{n,n} - \zeta \E[\ln {\cal R}_{n,n}] \\
%  &\quad-\frac{- ne+\zeta \ln(n)+n(1-\alpha)+e R_{n,n(1-\alpha)} - \zeta \E[\ln {\cal R}_{n,n(1-\alpha)}]- 0.7 }{1-\pi_{n,n(1-\alpha)}(n)} - ne+\zeta \ln(n)  \\
%  &= -n(e-1)+eR_{n,n} + \zeta( \ln(n)- \E[\ln {\cal R}_{n,n}]) \\
%  &\quad-\frac{- n(e-1)+e R_{n,n(1-\alpha)}+\zeta ( \ln(n) - \E[\ln {\cal R}_{n,n(1-\alpha)}])-n\alpha- 0.7 }{1-\pi_{n,n(1-\alpha)}(n)}. \label{eq:a15k}\end{align}
In the last passage above we get term (\ref{eq:akkk}) that is of the order $\mathcal{O}(n^{-1})$. In fact, in Lemma \ref{th:minorazione_errore} we have lower bounded function $ \overline{f(i)}$ that includes term $\E[\ln {\cal R}_{n,n}]$. By
result (\ref{eq:effe}) we have \be \ln(n)- \E[\ln {\cal R}_{n,n}] \le -\ln(1-e^{-1})
+\frac{1}{(e-1)}n^{-1} + \mathcal{O}(n^{-2}). \label{eq:effeq}\ee
 Using Property 3b
%\[ e R_{n,n} = n(e-1) - 0.5 -\frac{7}{24}n^{-1}+ \mathcal{O}(n^{-2})\]
and Jensen's inequality $\E[\ln {\cal R}_{n,n(1-\alpha)}] \le \ln(R_{n,n(1-\alpha)})$, term (\ref{eq:akkk}) becomes
\be  -0.5 -\zeta\ln(1-e^{-1})+ \left(\frac{\zeta}{(e-1)}-\frac{7}{24}\right)n^{-1}
+ \mathcal{O}(n^{-2})=  \mathcal{O}(n^{-1}), \label{eq:a16k}\ee
having exploited the relation  $-0.5 -\zeta\ln(1-e^{-1})=0$. As for term (\ref{eq:a15k}), we use expansions
\[ e R_{n,n(1-\alpha)} = n(e -e^{\frac{\alpha }{\alpha -1}}) + e^{\frac{\alpha}{\alpha-1 }} \frac{(2 \alpha -1)}{2 (\alpha -1)^2}+\mathcal{O}(n^{-1})\]
\be \ln(R_{n,n(1-\alpha)}) = \ln(n) + \ln\left(1-e^{\frac{1}{\alpha -1}} \right)- \frac{
e^{\frac{1}{\alpha-1 }}(2 \alpha -1)}{2 ( e^{\frac{1}{\alpha-1 }}-1) (\alpha -1)^2}n^{-1}
+\mathcal{O}(n^{-2})
 \ee
and  inequality (\ref{eq:akkk}) (\ref{eq:a15k}) becomes
 \be \Delta L(n,-\alpha n) \le  - \frac{s(n,\alpha)}{1-\pi_{n,n(1-\alpha)}(n)}+ \mathcal{O}(n^{-1}), \label{eq:a16k}\ee
 where
 \be s(n,\alpha)=n(1-\alpha-e^{\frac{\alpha }{\alpha -1}}) + e^{\frac{\alpha}{\alpha-1 }}
\frac{(2 \alpha -1)}{2 (\alpha -1)^2}-\zeta\ln\left(1-e^{\frac{1}{\alpha -1}} \right)- 0.7.
\label{eq:a21z} \ee

Function $s(n,\alpha)$ is negative only in a small interval beyond $\alpha=0$. It crosses the axis
at $\alpha_0$ that we find by expanding $s(n,\alpha)$ around $\alpha=0$ up to the second power and
for large $n$ . We get \be s(n,\alpha)= n\frac{\alpha^2 }{2}-0.7 + \mathcal{O}(n\alpha^3), \ee
which shows that $s(n,\alpha)$ switches from negative to positive at about
\[ \alpha_0= \sqrt{1.4/n}+o(1/\sqrt{n}), \]
and then remains positive up to $\alpha=(n-1)/n$. Therefore, from (\ref{eq:a16k}) we see that an
$n_0$ exists such that for $\alpha_0 < \alpha \le 1$ and
 all $n> n_0$ we have $\Delta L(n,n(1-\alpha))<0$.

\textbf{Part Ib}. Here we consider the case $h/n=\alpha \le \alpha_0$, i.e., $h/n \le
\alpha_0=\sqrt{1.4/n}+o(1/\sqrt{n})$. This means that we have 
\be h \le \sqrt{1.4 n}+o(1/\sqrt{n}). \label{eq:a121}  \ee

 The (\ref{eq:a14o}) can be bounded as follows 
  \begin{align} \Delta L(n,-h)&\le n+\sum_{i=2}^{n} \pi_{n,n}(i)
L(i)-\left( n-h+\sum_{i=2}^{n} \pi_{n,n-h}(i) L(i)\right)+ \pi_{n,n-h}(n) L(n)\nonumber \\
 &= h+\sum_{i=2}^{n} (\pi_{n,n}(i)-\pi_{n,n-h}(i)) L(i) + \mathcal{O}(c^{n}).
\end{align}
By substituting the expression $L(i)=ie-\varepsilon(i)$ (Theorem \ref{th:elle1}), condition (\ref{eq:a3}) turns into
 \be \overline{\varepsilon(i)} =\sum_{i=2}^{n}\varepsilon(i) (\pi_{n,n-h}(i)-\pi_{n,n}(i)) <
e(R_{n,n-h}-R_{n,n}) - h - \mathcal{O}(c^{n}). \label{eq:a14c} \ee
We use the expansion
\be R_{n,n-h}-R_{n,n}=he^{-1}\left(1+\frac{h+1}{2}n^{-1} +\frac{4h^2+18h+7}{24}n^{-2}
\right)+\mathcal{O}(n^{-3}). \label{eq:exp20} \ee
Furthermore, since $\pi_{n,n-h}$ statistically dominates $\pi_{n,n}$ (Lemma \ref{th:dominance1}),
we use Corollary \ref{th:corol2}, to show that $ \overline{\varepsilon(i)} \le \overline{l(i)} $,
where we have adopted the new distributions. Therefore, from (\ref{eq:a14c}), the thesis is
proved true by showing that
 \be \overline{\varepsilon(i)} \le \overline{l(i)} = \sum_{i=2}^{n}l(i) (\pi_{n,n-h}(i)-\pi_{n,n}(i))<
\frac{(h+1)h}{2}n^{-1} +\frac{(4h^2+18h+7)h}{24}n^{-2} + \mathcal{O}(n^{-3}).\label{eq:aa15} \ee

We prove the above inequality by bounding $\overline{l(i)} $ exactly as we bounded $\overline{g(i)} $ in
Lemma \ref{th:min_Delta_errore}. Actually, this evaluation is simpler, as function $l(n)$
coincides with the first part of $g(n)$. We take the power series at $a=R_{n,n-h}$, and get the
corresponding of (\ref{eq:kk3}), where now the moments are evaluated according to the
distributions in (\ref{eq:aa15}). We then substitute the asymptotic expansions to
%  \begin{align}
%  \frac{R_{n, n-h}-R_{n,n}}{R_{n,n}} = \frac{h}{e-1}n^{-1} & +\mathcal{O}(n^{-2}) \label{eq:exp21}\\
% - \frac{E[({\cal R}_{n,n-h}-R_{n,n})^2]}{2R_{n,n}'^2}+\frac{E[({\cal
% R}_{n,n}-R_{n,n})^2]}{2R_{n,n}'^2}&=\mathcal{O}(n^{-2}), \label{eq:exp22}\\
% ....&=\mathcal{O}(n^{-2}),
%  \end{align}
 get
 \be \overline{l(i)} \le \nu \frac{h}{e-1}n^{-1}+ \mathcal{O}(n^{-2}),
 \label{eq:a17} \ee
then (\ref{eq:aa15}) is true if the following is true
 \be \frac{\nu }{e-1}n^{-1} <
\frac{h+1}{2}n^{-1} + \mathcal{O}(n^{-2}). \label{eq:a18} \ee
Disregarding term $\mathcal{O}(n^{-2})$, the above inequality is always verified as by Lemma
\ref{th:min_Delta_errore} we have $\nu < e-1$. Under this hypothesis we can always find a finite
$n_1$ such that for all $n>n_1$ inequality (\ref{eq:a18}) holds.

 \textbf{Part II.} The (\ref{eq:a14o}) becomes
\be  \Delta L(n,k)=-k+\sum_{i=2}^{n}(\pi_{n,n}(i)-\pi_{n,n+k}(i))L(i) + \mathcal{O}(c^{n}). \label{eq:a132}
\ee Substituting $L(i)=ie-\varepsilon(i)$, condition (\ref{eq:a3}) transforms into
 \be \sum_{i=2}^{n}\varepsilon(i)
(\pi_{n,n}(i)-\pi_{n,n+k}(i)) \ge e(R_{n,n}-R_{n,n+k}) -k + \mathcal{O}(c^{n}). \label{eq:a132}
\ee
If we disregard term $\mathcal{O}(c^{n})$ the inequality above is always verified for any $k>0$.
In fact, by Property \ref{pr 4}b in Appendix B, the right hand term is negative. On the other side, by
Lemmas \ref{th:max_Delta_errore}, \ref{th:dominance}, \ref{th:dominance1}, and property
(\ref{eq:app01}), the left hand side cannot be negative (actually we can show it is positive). Therefore, we
can find an $n_2$, independent of $k$, even when $k \rightarrow \infty$, such that the above
inequality is satisfied for any $n>n_2$.

Then we numerically show that (\ref{eq:a3}) holds up to $\max[n_0,n_1,n_2]$, and the whole theorem is proved.
\end{IEEEproof}

\section{Conclusions} In this paper we have theoretically proved results about the Frame Aloha protocol that up to now were only numerically verified. In particular we have shown that the strategy that minimizes the time to the identification of a known number of tags is the one that sets at each frame the frame length $r$ equal to the backlog $n$. Furthermore we have shown that the optimal asymptotical  efficiency is $e^{-1}$, and derived tight upper and lower bounds to the identification time.

\appendices
\section{}\label{sec:appendix_A}

%{\bf \large Appendix}

We make use of the concept of stochastic dominance of first order. Given two non negative RV $X$ and $Y$, the probability distribution $\pi_X$ of $X$ is said to stochastically
dominate $\pi_Y$ of $Y$ if their cumulative distributions $F_X$ and $F_Y$ are such that
\[ F_X(i) \le F_Y(i), \qquad \forall i. \]
If the property above holds true, and $g(i)$ is a weakly increasing function, then the following
property holds \cite{fishburn}:
 \be \sum_i g(i)(\pi_X(i)-\pi_Y(i)) \ge 0. \label{eq:app01}\ee

\begin{lemma}\label{th:dominance}
if $\pi_X$ statistically dominates $\pi_Y$, and if $u(i)-u(i-1) \ge h(i)-h(i-1), \forall i$,
we have
 \be \sum_i u(i)(\pi_X(i)-\pi_Y(i)) \ge \sum_i h(i)(\pi_X(i)-\pi_Y(i)). \label{eq:app02}\ee
\end{lemma}
\begin{IEEEproof} in fact, (\ref{eq:app02}) holds if the following relation holds
 \be \sum_i (u(i)-h(i))(\pi_X(i)-\pi_Y(i))\ge 0. \label{eq:app02b}\ee
The above is true if $u(i)-h(i)$ is weakly increasing, which holds by hypothesis.
\end{IEEEproof}

If $u(x)$ and $h(x)$ are defined over the real interval the comprises all the values of RVs $X$
and $Y$, since
\[ \frac{d}{dx}(u(x)-h(x)) \ge 0, \quad \forall x \]
is a sufficient condition for $u(i)-h(i)\ge 0, \forall i$, we have
\begin{corollario}\label{th:dominancex}
if $\pi_X$ statistically dominates $\pi_Y$, and if $du(x)/dx \ge dh(x)/dx, \forall x$, then
inequality (\ref{eq:app02b}) holds.
\end{corollario}

With the notation used in the paper we have:
\begin{lemma}\label{th:dominance1}
\begin{enumerate}
 \item distribution $\pi_{n+1,r}$ stochastically dominates, in the first order, $\pi_{n,r}$;
 \item $\pi_{n+1,r+1}$ stochastically dominates, in the first order, $\pi_{n,r}$;
 \item $\pi_{n,r}$ stochastically dominates, in the first order, $\pi_{n,r+1}$.
\end{enumerate}
\end{lemma}
\begin{IEEEproof}
denoted by $F_{n,r}(i)=\PR({\cal R}_{n,r} \le i)$ the cumulative distribution function of ${\cal
R}_{n,r}$, 

1) we must  show that \be F_{n+1,r}(i) \le F_{n,r}(i),\qquad \forall i. \ee

The experiment that provides ${\cal R}_{n+1,r}$ can be composed of two subsequent
experiments: the first is the experiment that provides ${\cal R}_{n,r}$, and the second experiment
adds to the frame the $(n+1)$-th tag, which can be either collided or not. Let ${\cal R}_1$ denote
the increase in the number of collided tags the second experiment causes, either $0$, $1$, or $2$.
Therefore we have $ {\cal R}_{n+1,r}= {\cal R}_{n,r}+{\cal R}_1$, and also
\[ \PR({\cal R}_{n+1,r} >i)= \PR( {\cal R}_{n,r}+{\cal R}_1 >i)= \PR( {\cal R}_{n,r}>i)+
 \sum_{k \ge 0}\PR( {\cal R}_{n,r} = i-k)\PR({\cal R}_1 > k) > \PR( {\cal R}_{n,r}>i), \]
which prove the the first point.

2) Let us pick a slot, say slot $r+1$, in the frame of length $r+1$ with $n+1$
tags. We denote by $X$ the number of tags in slot $r+1$. Therefore, we have
\[ F_{n+1,r+1}(i|X=0) = F_{n+1,r}(i), \]
\[ F_{n+1,r+1}(i|X=1) = F_{n,r}(i), \]
\[ F_{n+1,r+1}(i|X>1) = \frac{\PR(X>1|{\cal R}_{n+1,r+1} \le i)}{\PR(X>1)}F_{n+1,r+1}(i). \]
Then we can write
\[ F_{n+1,r+1}(i)= \PR(X=0) F_{n+1,r}(i) + \PR(X=1) F_{n,r}(i) + \PR(X>1|{\cal R}_{n+1,r+1} \le
i)F_{n+1,r+1}(i), \]
 and by the result at first point we have
\[ F_{n+1,r+1}(i) \le \PR(X=0) F_{n,r}(i) + \PR(X=1) F_{n,r}(i) +
\PR(X>1|{\cal R}_{n+1,r+1} \le i)F_{n+1,r+1}(i), \] which provides
\[ F_{n+1,r+1}(i) \le \frac{1-\PR(X>1)}{1-\PR(X>1|{\cal R}_{n+1,r+1} \le i)} F_{n,r}(i). \]

Since $\PR({\cal R}_{n+1,r+1} \le n+1)=1$, to prove the thesis we must show that
 \[ \PR(X>1|{\cal R}_{n+1,r+1} \le i) \le \PR(X>1|{\cal R}_{n+1,r+1} \le n+1)= \PR(X>1),
\qquad i\le n. \] The above is proven also if we prove the more stringent condition
 \be \PR(X>1|{\cal R}_{n+1,r+1} =j) \le \PR(X>1|{\cal R}_{n+1,r+1} =j+1), \qquad
2 \le j \le n. \label{eq:lamma21}\ee We have
 \be \PR(X>1|{\cal R}_{n+1,r+1} =j+1)=\sum_c
\frac{c}{r+1}\PR({\cal C}_{n+1,r+1}=c|{\cal R}_{n+1,r+1} =j+1), \label{eq:lamma22}\ee where ${\cal
C}_{n+1,r+1}$ represents the number of collided slots.

The experiment that provides collided slots starting from $j+1$ collided tags is the one that
distributes collided tags among $n+1$ slots; then tags that belong to successful slots are
re-distributed among the $n+1$ slots and the procedure is repeated until no more successful slots
are present.

The experiment carried out starting with $j$ collided tags can be derived from the experiment with
$j+1$ collided tags in the following way. From the $j+1$ collided tags of the former experiment we
disregard one tag at random. Then we have two cases; either the number of collided slots is
unchanged, or a collided slot (with two tags) is transformed into a successful slot, whose tag is
re-assigned, so that the number of collided slots is decreased by one. This proves that we have
 \[ \PR({\cal C}_{n+1,r+1}=c|{\cal R}_{n+1,r+1} =j)\le \PR({\cal C}_{n+1,r+1}=c|{\cal R}_{n+1,r+1}
 =j+1), \qquad \forall c, 2 \le j \le n \] which, replaced into (\ref{eq:lamma22}), proves
 (\ref{eq:lamma21}) and the thesis is proved.

3) Much like the previous case we have
\[ F_{n+1,r+1}(i|X=0) = F_{n+1,r}(i), \]
\[ F_{n+1,r+1}(i|X>0) = \frac{\PR(X>0|{\cal R}_{n+1,r+1} \le i)}{\PR(X>0)}F_{n+1,r+1}(i), \]
and
\[ F_{n+1,r+1}(i)= \PR(X=0) F_{n+1,r}(i) + \PR(X>0|{\cal R}_{n+1,r+1} \le i)F_{n+1,r+1}(i), \]
which provides
\[ F_{n+1,r+1}(i) = \frac{\PR(X=0)}{\PR(X=0|{\cal R}_{n+1,r+1} \le i)} F_{n+1,r}(i). \]
To prove the thesis, i.e., $F_{n+1,r+1}(i) \ge F_{n+1,r}(i) , \forall i$, we must show that
 \[ \PR(X=0|{\cal R}_{n+1,r+1} \le i) \ge \PR(X=0|{\cal R}_{n+1,r+1} \le n+1)= \PR(X=0),
\qquad 2 \le i\le n, \] or the more stringent condition
 \be \PR(X=0|{\cal R}_{n+1,r+1} =j) \ge \PR(X=0|{\cal R}_{n+1,r+1} =j+1), \qquad
2 \le j \le n. \label{eq:lemma21}\ee

Inequality (\ref{eq:lemma21}) is equal to (\ref{eq:lamma21}) but the facts that the inequality sign is reversed and  $X>1$ is replaced by
$X=0$ (empty slot). The proof proceeds exactly as in the former case, replacing collided slots
with empty ones. Now, when in the experiment we discard one tag, the number of empty slots, ${\cal
E}_{n+1,r+1}$, either remains the same or increases by one. This shows that
 \be \PR({\cal E}_{n+1,r+1}=\varepsilon|{\cal R}_{n+1,r+1} =j)\ge \PR({\cal E}_{n+1,r+1}=\varepsilon|{\cal R}_{n+1,r+1} =j+1),
 \label{eq:lam23} \ee
and the thesis is proven.
\end{IEEEproof}

\section{}\label{sec:appendix_B}

A recursive expression of $p_{n, r_n}(s,c)$, the probability of having $s$ successes and $c$
collided slots in the first frame, is given in \cite{schoute}. A closed-form
expression for $p_{n, r}(s)$ is given in the following

\begin{property}\label{th:distrib} the distribution $p_{n,r}$ is given by
 \be p_{n,r}(i)=\sum_{k=i}^{m} (-1)^{k+i} {k \choose i}
X_{n,r}(k),\qquad 0\le i\le m, \label{eq: distri01} \ee
where $m=\min\{n,r\}$, and
\begin{equation}
X_{n,r}(k)= {r \choose k} \frac{n!}{(n-k)!}\left(\frac{1}{r}\right)^k
\left(\frac{r-k}{r}\right)^{n-k}, \label{eq: 130}
\end{equation}
with $k\le m$.
\end{property}
Furthermore we have
 \be p_{n+1,r+1}(i+1) = p_{n,r}(i)
\frac{n+1}{i+1}\left(\frac{r}{r+1}\right)^{n},\qquad 0\le i\le m. \label{eq:recur1}\ee
%{\bf Theorem 1}:
\begin{IEEEproof}
Let $A_1, A_2, \ldots, A_r$ be $r$ non-disjoint events. The probability that exactly $t$ among
these events jointly occur is given by \cite{feller}
\begin{equation}
P_t = X_t-{t+1 \choose t} X_{t+1} + {t+2 \choose t} X_{t+2} - \ldots +(-1)^{r-t} {r \choose t} X_r
\label{eq: 126}
\end{equation}
where
\begin{align}
X_1 & = \displaystyle \sum \PR(A_i) \nonumber\\
X_2 & = \displaystyle \sum_{i\neq j} \PR(A_i A_j) \nonumber\\
X_3 & = \displaystyle \sum_{i\neq j\neq k} \PR(A_i A_j A_k) \label{eq: 125}
\end{align}
and so on. Summations involve all possible combinations in such a way that each $n$-string appears
just once, and the number of the terms $X_k$ is ${r \choose k}$.

In our case the event $A_i$ is defined as the occurrence of just one transmission, out of $n$, in
slot $i$ of a frame composed of $r$ slots, and the probability of any of the $k$-string is given by

\[ \PR(A_{j_1}A_{j_2}\ldots A_{j_k})=\frac{n!}{(n-k)!}\left(\frac{1}{r}\right)^k
\left(\frac{r-k}{r}\right)^{n-k},
\]
$k\le m$, which by (\ref{eq: 126}) and (\ref{eq: 125}) proves the first part of the theorem.

The proof of the second part comes from (\ref{eq: distri01}) and (\ref{eq: 130}), observing that
\[
X_{n+1,r+1}(k+1)=\frac{n+1}{k+1}\left(\frac{r}{r+1}\right)^{n}X_{n,r}(k).
\]
and rearranging terms.
\end{IEEEproof}

\begin{property}\label{lem:expo_pi}
the sequence $\{\pi_n(n)\}_n$ for $n > 16$ is bounded as:
 \be\label{eq:ub_pi_n}
\pi_n(n)<\bar{\pi}_n(n)=3.47\cdot 10^{-3}\cdot 0.9157^n+59.79\cdot 0.4157^n.
%\frac{9}{1168}\left[(73-3\sqrt{73})\left(\frac{3+\sqrt{73}}{12}\right)^n+(73+3\sqrt{73})\left(-\frac{3-\sqrt{73}}{12}\right)^n\right].
\ee
\end{property}
\begin{IEEEproof} once the number of tags that participates in a frame is fixed, adding a slot to the frame
decreases the probability of having no successes, or, in other terms, \be
\pi_{n+1}(n+1)<\pi_{n+1,n}(n) \label{eq: ineq0}.\ee On the other side, considering the outcome of
the $(n+1)$-th tag being added to the frame, we can write
 \be \pi_{n+1,n}(n)=\pi_n(n)\PR(X=1)+\pi_n(n-1)\PR(X=2), \label{eq:ineq_pi_n} \ee
 where $X$ denotes the increase in
the collided tags caused by the $(n+1)$-th tag. In the case represented by $\pi_n(n-1)$ there is
only one success that the new added tag turns into two more collisions, and this happens with
probability
\[ \PR(X=2)=1/n. \]
In the case represented by $\pi_n(n)$ the added tag must select one of the collided slots, and
this happens with probability
\[ \PR(X=1)=\frac{\E[{\cal C}_n|{\cal S}_n=0]}{n}<\frac{n}{2}\cdot\frac{1}{n}=\frac{1}{2}. \]
From recursion (\ref{eq:recur1}) one has \be \pi_{n}(n-1)=p_{n}(1)=p_{n-1}(0)\cdot
S_n=\pi_{n-1}(n-1)\cdot S_n. \label{eq: ineq1} \ee
Using (\ref{eq: ineq0}), (\ref{eq:ineq_pi_n}) and (\ref{eq: ineq1}) we finally get
\[
\pi_{n+1}(n+1)<\frac{1}{2}\pi_n(n)+\frac{S_{n}}{n}\pi_{n-1}(n-1),
\]
and, taking advantage of Property \ref{th:properties}b, we may write
\[
\pi_{n+1}(n+1)<\frac{1}{2}\pi_n(n)+\frac{S_{15}}{15}\pi_{n-1}(n-1),
\]
for $n\ge 15$. This means that it is possible to build a sequence $\{\bar{\pi}_n(n)\}$, that upper
bounds the actual sequence $\{\pi_n(n)\}$, through the recurrence
\[
\bar{\pi}_{n+1}(n+1)=0.5\ \bar{\pi}_n(n)+0.381\ \bar{\pi}_{n-1}(n-1),
\]
for $n\ge 16$, with initial conditions $\bar{\pi}_{14}(14)=\pi_{14}(14)\approx 1.285\cdot 10^{-3}$
and $\bar{\pi}_{15}(15)=\pi_{15}(15)\approx 8.106\cdot 10^{-4}$. The solution of the above
difference equation is \be\label{eq:sol_diff_eq} \bar{\pi}_n(n)=3.47\cdot 10^{-3}\cdot
0.9157^n+59.79\cdot (-0.4157)^n,
%\frac{9}{1168}\left[(73-3\sqrt{73})\left(\frac{3+\sqrt{73}}{12}\right)^n+(73+3\sqrt{73})\left(\frac{3-\sqrt{73}}{12}\right)^n\right],
\ee for $n\ge 16$. From this, bound (\ref{eq:ub_pi_n}) is immediate.
\end{IEEEproof}

In the analysis carried out in the paper we make use of properties of RVs ${\cal S}_n$ and ${\cal
 R}_n=n-{\cal S}_n$, listed below, that can be proved with standard tools.
\begin{property}\label{th:properties}
\end{property}
\begin{enumerate}
 \item[a)] $ \displaystyle \E[{\cal S}_{n,r}]=S_{n,r}= n\left(1-\frac{1}{r}\right)^{n-1};
 $ %\label{eq: th21}\ee
 \item[b)] $ R_n/n$ is an increasing function of $n$ such that
 $\frac{\displaystyle R_{n}}{n}=1-e^{-1}-\frac{e^{-1}}{2n}-\frac{7e^{-1} }{24n^2}-\frac{3e^{-1}}{16
n^3}+\mathcal{O}(n^{-4})$;
\item[c)] $R_{n+1}-R_{n}$ is a decreasing function of $n$ with
 $(R_{n+1}-R_{n})=1-e^{-1} +\frac{7e^{-1} }{24n^2}+\mathcal{O}(n^{-3})$.
\end{enumerate}

\begin{property}\label{pr 4} \end{property}
\begin{enumerate}
 \item[a)] $ R_{n,n+k-1}-R_{n,n+k}< e^{-1} $, for all $n\ge 1$ and $k\ge1$;
\item[b)] $ R_{n,n}-R_{n,n+k}< k e^{-1} $, for all $n\ge 1$ and $k\ge1$.
\end{enumerate}

\begin{IEEEproof} for the first point it is
\[
R_{n,n+k-1}-R_{n,n+k}= S_{n,n+k}-S_{n,n+k-1} \le \max_{k\in \{1,2,\ldots\} }\
(S_{n,n+k}-S_{n,n+k-1})\le \sup_{k\in [1,\infty)}\ \frac{\partial }{\partial k}S_{n,n+k}
\]
for all $n\ge 1$. The derivative with respect to $k$ is
\begin{equation}\label{eq:deriv}
 \frac{\partial }{\partial k}S_{n,n+k} = \frac{n}{n+k} \frac{S_{n-1,n+k}}{n+k},
\end{equation}
where $S_{n-1,n+k}/(n+k)$, the throughput per slot, is a decreasing function of $k$, for $k\ge 1$.
This means that also (\ref{eq:deriv}) is a decreasing function of $k$, and therefore the maximum
is achieved for $k=1$:
\[
 \frac{\partial }{\partial k}S_{n,n+k}\le \frac{n}{(n+1)^2} S_{n-1,n+1} = \frac{n(n-1)}{(n+1)^2}\left(1-\frac{1}{n+1}\right)^{n-2}< e^{-1}.
\]

Point 4b comes straightforwardly from point 4a.
\end{IEEEproof}

\begin{property}\label{th:moments}
here we show how to derive moments of variable ${\cal S}_{n,r}$. Moments
for variable ${\cal R}_{n,r}$ can be derived by the relation ${\cal R}_{n,r}=n-{\cal S}_{n,r}$.
The first order moment is given above in Property \ref{th:properties}a. For the evaluation of
higher order moments we express ${\cal S}_{n,r}$ as the sum of binary variables $X_{n,r}(i)$,
where $X_{n,r}(i)$ takes value $1$ if in the corresponding $i$-th slot of the frame there is only
one tag, i.e., a success. Hence
\[ \E[{\cal S}_{n,r}^k]=\E\left[\left(\sum_{i=1}^r X_{n,r}(i)\right)^k\right]. \]
We use the multinomial theorem, that gives
\[ \E\left[\left(\sum_{i=1}^r X_{n,r}(i)\right)^k\right]= \sum_{k_1+k_2+\ldots+k_r=k} \frac{k!}{k_1!k_2!\ldots k_r!}
\E\left[\prod_{j=1}^r X_{n,r}^{k_j}(j)\right]. \]
We have
\[\E\left[\prod_{j=1}^r X_{n,r}^{k_j}(j)\right]=\sum \PR(X_{n,r}(1)=x_1;X_{n,r}(2)=x_2;\ldots X_{n,r}(r)=x_r)\prod_{j \in \Omega} x_j^{k_j}, \]
where the summation is extended to the whole space of outcomes, whereas the product is extended
to indexes $j \in \Omega=\{j_1,j_2,\ldots,j_{|\Omega|}\}$ for which $k_j>0$, and $1\le |\Omega|\le \min \{r,k\}$. We consider the case where $k\le r$. This allow us to write
\begin{align*}
\E\left[\prod_{j=1}^r X_{n,r}^{k_j}(j)\right]&=\PR(X_{n,r}(j_1)=1;X_{n,r}(j_2)=1;\ldots X_{n,r}(j_{|\Omega|})=1)\\
&= \PR(X_{n,r}(j_1)=1) \cdot \prod_{t=2}^{|\Omega|} \PR(X_{n,r}(j_t)=1 | X_{n,r}(j_{t-1})=1, X_{n,r}(j_{t-2})=1,\ldots, X_{n,r}(j_1)=1)\\
&= \prod_{t=1}^{|\Omega|} \PR(X_{n-t+1,r-t+1}(j_t)=1) = \prod_{t=1}^{|\Omega|} \frac{S_{n-t+1,r-t+1}}{r-t+1},
\end{align*}
where we have used the chain rule for probabilities and the fact that knowing the outcomes of some slots reduces the problem.
Furthermore, sequences $k_1,k_2,\ldots k_n$ that represent permutations of a single sequence provide the same $\E[\prod_{j=1}^r X_{n,r}^{k_j}(j)]$. As a consequence, in the case where $k\le r$, the $k$-th moment of RV ${\cal S}_{n,r}$ can be written as
\[
 \E[{\cal S}_{n,r}^k]= \sum_{i=1}^k a_i \prod_{t=1}^i \frac{S_{n-t+1,r-t+1}}{r-t+1} = \frac{S_{n,r}}{r}\left(a_1+ \frac{S_{n-1,r-1}}{r-1}\left(a_2+\ldots \left(a_{k-1}+\frac{S_{n-k+1,r-k+1}}{r-k+1}a_k\right)\right)\right),
\]
where $a_i$ is the number of combinations where $|\Omega|=i$.

For example, for $k=2\le r$, we have $r$ terms corresponding to $k_j=2$, being all the others zero ($|\Omega|=1$ and $a_1=r$), and $r(r-1)$ terms of type $k_i=1,k_j=1$, $i\ne j$, being all the others zero ($|\Omega|=2$ and $a_2 = r(r-1)$). This provides
\[ \E[{\cal S}^2_{n,r}]=S_{n,r}+S_{n,r}S_{n-1,r-1}. \]
In a similar way we have found
\begin{align*} \E[{\cal S}_{n,r}^3]&=S_{n,r}+ 3S_{n,r}S_{n-1,r-1}+ S_{n,r}S_{n-1,r-1}S_{n-2,r-2}, \\
 \E[{\cal S}_{n,r}^4]&=S_{n,r}+ 7S_{n,r}S_{n-1,r-1}+ 6S_{n,r}S_{n-1,r-1}S_{n-2,r-2}+ S_{n,r}S_{n-1,r-1}S_{n-2,r-2}S_{n-3,r-3}, \\
 \E[{\cal S}_{n,r}^5]&=S_{n,r}+ 15S_{n,r}S_{n-1,r-1}+ 25S_{n,r}S_{n-1,r-1}S_{n-2,r-2}+ 10 S_{n,r}S_{n-1,r-1}S_{n-2,r-2}S_{n-3,r-3} \\
 &\quad+S_{n,r}S_{n-1,r-1}S_{n-2,r-2}S_{n-3,r-3}S_{n-4,r-4} .
 \end{align*}
\end{property}

\bibliography{biblio}
\bibliographystyle{IEEEtran}

\end{document}